\documentclass[12pt]{article}

\usepackage{times}

\usepackage{multirow}

\usepackage{graphicx}
\usepackage{caption}
\usepackage{subcaption}

\newenvironment{sciabstract}{%
\begin{quote} \bf}
{\end{quote}}

\title{Disinformation and Misinformation on Twitter during the Novel Coronavirus Outbreak}

\author
{Binxuan Huang, Kathleen M. Carley\\
\\
\normalsize{School of Computer Science, Carnegie Mellon University,}\\
\normalsize{5000 Forbes Avenue, Pittsburgh, PA 15213, USA}\\
\\
\normalsize{E-mail:  binxuanh@cs.cmu.edu}
}

\date{}

\begin{document} 




\maketitle


\begin{sciabstract}
 As the novel coronavirus spread globally, a growing public panic was expressed over the internet. We examine the public discussion concerning COVID-19 on Twitter. We use a dataset of 67 million tweets from 12 million users collected between January 29, 2020 and March 4, 2020. We categorize users based on their home countries, social identities, and political orientation. We find that news media, government officials, and individual news reporters posted a majority of influential tweets, while the most influential ones are still written by regular users. Tweets mentioning ``fake news'' URLs and disinformation story-lines are also more likely to be spread by regular users. Unlike real news and normal tweets, tweets containing URLs pointing to ``fake news'' sites are most likely to be retweeted within the source country and so are less likely to spread internationally.
\end{sciabstract}


\section*{Introduction}

As 2020 began, an outbreak of a new respiratory disease that would come to be known as COVID-19 occurred. The disease was first reported from Wuhan, China on December 31, 2019 \footnote{https://www.who.int/emergencies/diseases/novel-coronavirus-2019}. On January 30, 2020 the World Health Organization (WHO) declared the outbreak a public health emergency of international concern. By May 7, 2020, more than 3,900,000 cases were confirmed worldwide spread across 214 countries and regions, and 270,057 people had died. Severe outbreaks has occurred in China, United States, and Europe.

As the novel coronavirus spread globally, a growing public panic was expressed over the internet. We tracked this panic using Twitter.  Twitter is one of major social media platforms where users expressed concerns about the outbreak of this disease, shared purported preventions and cures, discussed theories about where the disease came from,  and how governments were and should respond. A significant fraction of the information being shared was ``fake''  as noted by numerous news agencies reports\footnote{https://www.nytimes.com/2020/03/08/technology/coronavirus-misinformation-social-media.html?searchResultPosition=1}. Online fact checking sites, like Poynter\footnote{https://www.poynter.org/ifcn-covid-19-misinformation/}, put up new information each day about new disinformation stories. Our analysis of these stories\footnote{https://www.cmu.edu/ideas-social-cybersecurity/research/index.html} showed many types of disinformation stories: false preventions and cures, false claims about the nature of the disease, false diagnostic procedures, false origin stories, false emergency measures, false ``feel good'' stories, and so on.  

The research community has frequently turned to social media to study the spread of information, disinformation and misinformation \cite{allcott2017social,lazer2018science,lee2014facebook,babcock2020pretending}. Among the platforms studied are: Twitter \cite{vosoughi2018spread,shao2018anatomy,babcock2019different}, Facebook \cite{mocanu2015collective,bessi2015trend,del2016spreading}, and Youtube \cite{donzelli2018misinformation,qi2016misinformation,loeb2019dissemination}. Disinformation or ``fake news'' has recently draw attention primarily in a political context, such as studies around elections \cite{grinberg2019fake,allcott2017social,uyheng2019characterizing}. Bovet and Makse investigated 30 million tweets containing a link to news outlets preceding the election day. They found that the top influencers spreading traditional center and left leaning news largely influence the Clinton supporters, while top ``fake news''  spreaders influence Trump supporters \cite{bovet2019influence}. Grinberg et al. found that only small portion of individuals accounted for a majority of ``fake news''  sharing \cite{grinberg2019fake}. They also found conservative leaning users are more likely to engage with ``fake news''  sources. ``fake news''  also emerges in information about topics such as vaccination and natural disaster. Chiou and Tucker studied the role of social media in the dissemination of false news stories about vaccines. They documented that members of anti-vaccine Facebook groups tended to disseminate false stories beyond the group through diverse media. Gupta et al. in a study of fake images during Hurricane Sandy\cite{gupta2013faking} found that the top thirty users resulted in 90\% of retweets of fake images. Most prior research has focused on specific users, with little concern for the type of user or their geographic location.  An exception here is the work by Babcock and colleagues that shows that disinformation spread by celebrities or newsagencies has greater reach \cite{babcock2020pretending}, and Carley et al. \cite{carley2016high} that news agencies are typically the most retweeted users, particularly during disasters.

During a pandemic, trust in health authorities is critical to prevent the spread of the disease, to save lives, and to enable public safety. Misinformation is damaging and can even be deadly. Because of the severe consequence, it is critical to understand the spread of accurate and inaccurate information. A key problem in a global pandemic is that while these authorities, other than the World Health Organization are local, information and disinformation is spread globally.  Hence, disinformation from one country can undermine, even unintentionally, the heath authority in another country.  This may be particularly true when the information appears to come from a credible source such as a newsagency or government official. However, little is known about the spread of information, let alone misinformation, between countries. Little is known about the role of types of actors, such as news agencies, in the spread of information and disinformation particularly from a global perspective.
In this study, we examine the global spread of information related to key disinformation stories during the early stages of the global pandemic. We address four research questions:\\
1. What types of users send influential tweets in this global health emergency event? \\
2. Who is discussing disinformation stories? \\
3. Where in the world are those who discuss low credibility information? \\
4. What is the global network for discussing low credibility information?

\section*{Data and Definition}
\subsection*{Data collection}
To answer these questions, we monitored conversations about COVID-19 on Twitter starting from January 29, 2020 to March 4, 2020. We select a list of keywords to track Twitter's real-time conversations\footnote{https://developer.twitter.com/en/docs/tweets/filter-realtime/guides/basic-stream-parameters}. The list include ``coronavirus'', ``coronaravirus'', ``wuhan virus'', ``wuhanvirus'', ``2019nCoV'', ``NCoV'', ``NCoV2019''. There are 67.4 million tweets and 12.0 million users involved in this time period.  We recognize that most Twitter users are tweeting in English, and that much of the discussion around the coronavirus early on, used the English terms if it was on Twitter.  Hence, we used these predominantly English keywords, and the WHO terms.  Although this creates a bias toward the spread of English tweets, it does pick up a large number of tweets in other languages.

\begin{figure*}[!h]
    \centering
    \includegraphics[width=1\textwidth]{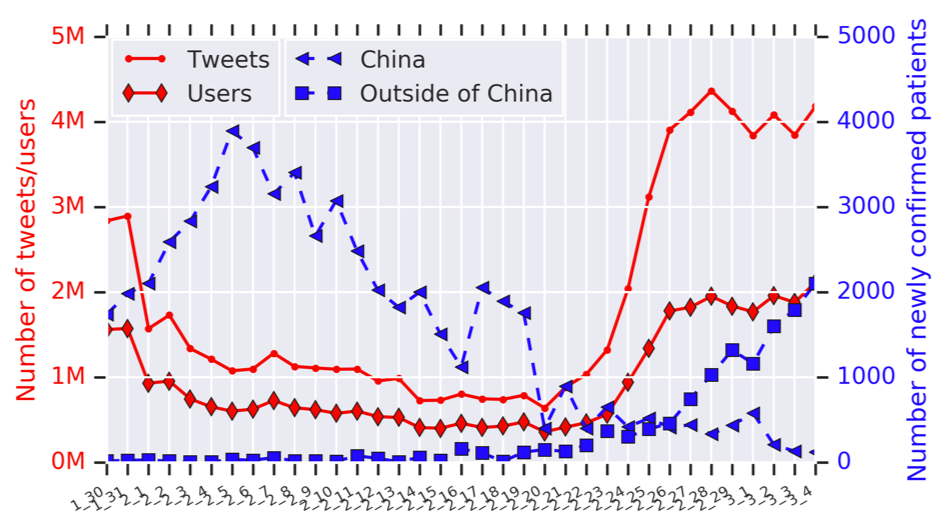}
    \caption{The red lines show the number of tweets and number of users each day. The blue lines represent the newly confirmed patients each day.}
    \label{ch6_trend}
\end{figure*}

In Fig. \ref{ch6_trend}, two red lines show the general trends for number of tweets and number of users involved each day. The two blue lines represent the number of newly confirmed patients each day in China as well as those outside of China reported by WHO \footnote{https://www.who.int/emergencies/diseases/novel-coronavirus-2019/situation-reports}. As shown in this figure, the volume of tweets first gradually reduced as the disease was contained inside China. As confirmed cases spread throughout the world, the volume of daily conversation in Twitter soared. By February 26, it was six times higher than on February 20 (3,904,293 v.s. 638,204).

To further determine users' location, social identity, and political orientation, we also collected the most recent 200 tweets and the following ties for each user who posted tweets between January 29, 2020 and March 4, 2020. Among 12,047,990 involved users, we successfully collected information for 11,951,739 users. These data are further fed into a state-of-the-art user profiling system \cite{huang2019hierarchical,huang2020identity}. Using this system we predict the users' home country based on these tweets with 92.96\% accuracy. We classify users' social identity into seven categories -- news media, news reporter, celebrity, government official, sport, company, and regular user. We also predict users' political orientation as liberal or conservative. We achieve accuracies of 95.4\% and 87.4\% on these standalone test datasets for identity classification and political orientation prediction respectively. Because our identity and political orientation classifiers are mainly trained on English users, we only apply these two classifiers to users whose major language is English.

\subsection*{Disinformation story-lines and news sites}
Following previous work \cite{lazer2018science} and \cite{grinberg2019fake}, we define ``fake news sites'' as ones that ``lack the news
media’s editorial norms and processes for ensuring the accuracy and credibility of information.'' We adopt three lists of ``fake news''  sites as proposed in \cite{grinberg2019fake}: black sites, orange sites, red sites. The black list contains a set of websites which published exclusively fabricated stories. The red list is a set of websites spreading falsehoods with a flawed editorial process. Sites labeled as orange represent cases where annotators were less certain that the falsehoods stemmed from a flawed editorial process. We further add a list of news sources as trusted news sites.
There are 20 black, 26 red, 25 orange ``fake news''  sites and 90 real news sites whose URLs appear in our collected data.

To study the conversation around specific  disinformation stories, we manually identified five disinformation story-lines. The first is a popular conspiracy that this novel coronavirus is a bio-weapon developed in a research lab. The remaining story-lines are about potential cures for this disease -- garlic, sesame oil, bleach, and chlorine dioxide. For each story-line, we retrieve tweets by searching for corresponding keywords in our COVID-19 corpus. For this body of tweets, we can say that they are discussing a particular disinformation story-line; however, at this point we cannot say whether or not the sender of the message is knowingly spreading disinformation, unwittingly spreading misinformation, joking about the story-line (satire), or pointing out that this story-line is not true and so countering disinformation.  What we can say is that they are taking part in the discussion around that story-line.

We show the number of tweets that contain one of the news URLs and a story-line in Supplementary Table \ref{fake_misinfo} for original tweets as well as retweets. As shown in Supplementary Table \ref{fake_misinfo}, most of source tweets with ``fake news'' URLs contain keywords related to the bio-weapon conspiracy. There is a high percentage of tweets both mentioning ``bio-weapon'' and ``fake news''  URLs compared to tweets mentioning ``bio-weapon'' and a trusted news source. This suggests that the bio-weapon disinformation came from these ``fake news'' sites.  In the case of retweets, a high percentage of retweets that mention red news or orange news sites also mention ``bio-weapon''. This suggests that these less credible news sites were critical in further spreading this conspiracy; particularly as tweets mentioning black news URLs and ``bio-weapon'' were less likely to be retweeted. It is likely that there are more tweets discussing these stories, than listed here, as selection on keyword tends to under-sample. 

\begin{table}[]
\caption{Number of source tweets and retweets with news URLs, overall and by story-line. }
\label{fake_misinfo}
\resizebox{\textwidth}{!}{
\begin{tabular}{|l|l|l|l|l|l|l|l|}
\hline
                                                                         & News & Total   & Bio-weapon       & Bleach     & Chlorine    & Garlic       & Sesame      \\ \hline
\multirow{4}{*}{\begin{tabular}[c]{@{}l@{}}Source \\ tweet\end{tabular}} & Black      & 3083    & 92 (2.98\%)      & 0 (0.00\%) & 0 (0.00\%)  & 0 (0.00\%)   & 0 (0.00\%)  \\ \cline{2-8} 
                                                                         & Red        & 53522   & 4,447 (8.31\%)   & 0 (0.00\%) & 0 (0.00\%)  & 5 (0.01\%)   & 0 (0.00\%)  \\ \cline{2-8} 
                                                                         & Orange     & 61162   & 4,879 (7.98\%)   & 0 (0.00\%) & 1 (0.00\%)  & 10 (0.02\%)  & 0 (0.00\%)  \\ \cline{2-8} 
                                                                         & Real       & 796071  & 1,245 (0.16\%)   & 0 (0.00\%) & 13 (0.00\%) & 169 (0.02\%) & 18 (0.00\%) \\ \hline
\multirow{4}{*}{Retweet}                                                 & Black      & 32302   & 73 (0.23\%)      & 0 (0.00\%) & 0 (0.00\%)  & 0 (0.00\%)   & 0 (0.00\%)  \\ \cline{2-8} 
                                                                         & Red        & 151249  & 15,467 (10.23\%) & 0 (0.00\%) & 0 (0.00\%)  & 2 (0.00\%)   & 0 (0.00\%)  \\ \cline{2-8} 
                                                                         & Orange     & 205362  & 24,916 (12.13\%) & 0 (0.00\%) & 0 (0.00\%)  & 11 (0.01\%)  & 0 (0.00\%)  \\ \cline{2-8} 
                                                                         & Real       & 2738551 & 4,291 (0.16\%)   & 0 (0.00\%) & 0 (0.00\%)  & 205 (0.01\%) & 25 (0.00\%) \\ \hline
\end{tabular}
}
\end{table}



\section*{Results}
\subsection*{What types of users spread influential tweets?}
There are 67,408,573 tweets posted by 12,047,990 users in total between January 29, 2020 and March 4, 2020. Among 5,448,250 English speaking users, 28.6\% exhibit bot-like characteristics and behavior. There are 96.44\% regular users (30.73\% are labeled as bots\footnote{We use bothunter with a 60\% cutoff which has a precision of .957 and a recall of .704. \cite{beskow2018introducing,beskow2020you,beskow2020social}}), 1.12\% news agencies (32.91\% of which are bots), 1.18\% news reporters (18.04\% of which are bots), and 1.05\% government officials (19.94\% of which are bots). Only 0.15\% of users are company accounts and the rest appear to be celebrities.  As companies and celebrities together make up less than 1\% of the users, we do not continue to analyze these groups in this paper.

\begin{figure*}[!h]
    \centering
    \includegraphics[width=\textwidth]{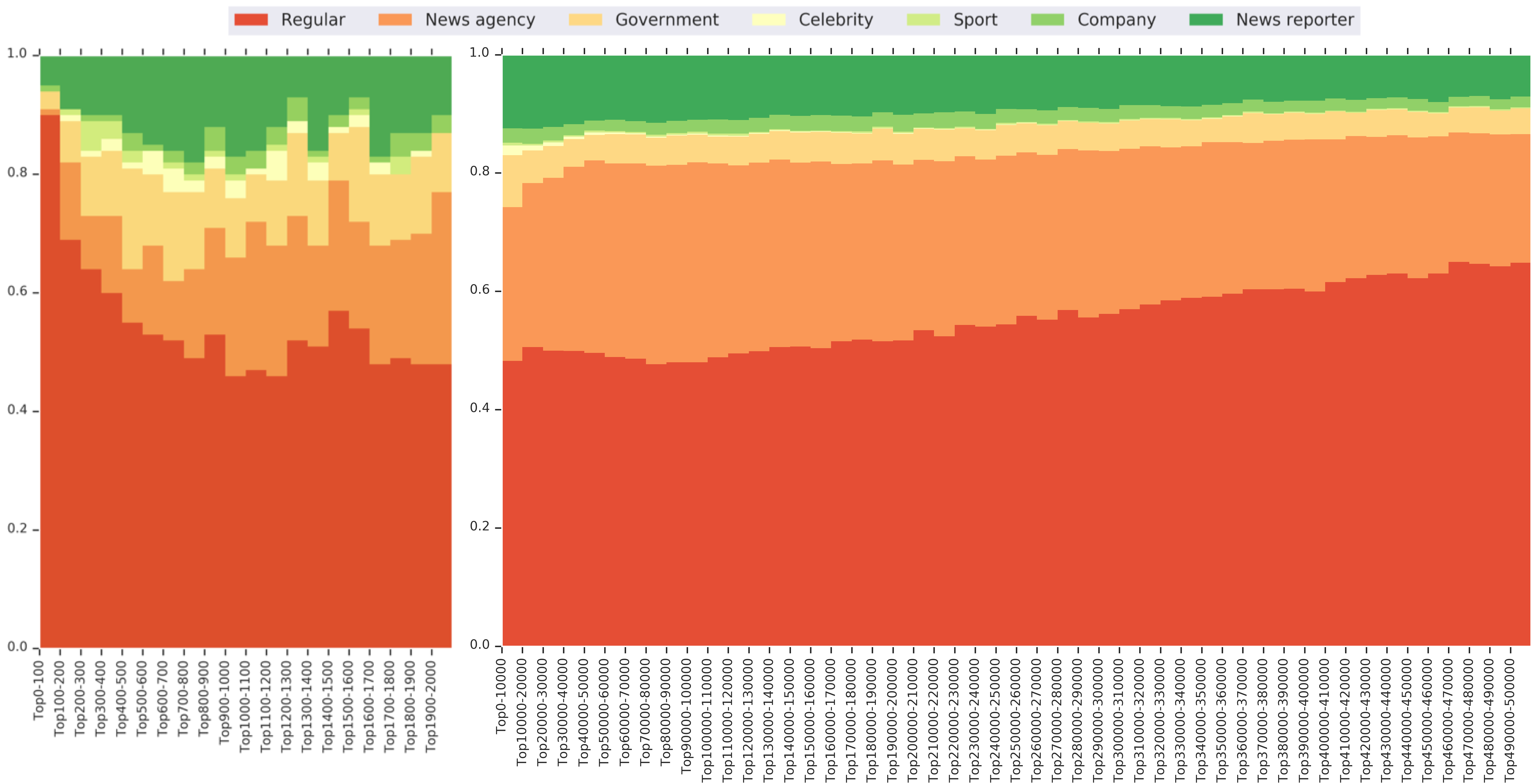}
    \caption{Two bar charts show identity distributions across top influential tweets. There are 20 bars in the left figure, each of which represents the identity distribution of source users of 100 tweeters. The leftmost bar is for the top 100 most retweeted tweets. From left to right, the number of retweets decrease. The right figure contains 50 bars. Each bar shows the identity distribution of 10000 tweeters. The leftmost bar is the identity distribution for the most retweeted 10000 tweets.}
    \label{ch6_top}
\end{figure*}

We define influential tweets as those which were the most retweeted. For these influential tweets, we identify the type of user (identity) mainly tweeting in English that posted it. In Fig. \ref{ch6_top}, two bar charts show the identity distribution across these most widely spread tweets. As expected, news medias play an important role in this event. 25.98\% and 12.38\% of top 10000 most influential tweets are posted by news agencies and individual news reporters. Government officials also contribute 8.79\% of these 10000 influential tweets. Even though less than 5\% of users are these official accounts, they contribute more than 50\% of these top 10000 tweets. Regular users posted 48.29\% of these influential tweets. The percentage is relatively small considering more than 95\% of the users are regular users. A deeper dive into these top 100 most influential tweets, shows that 90\% of the tweets are posted by regular users. Thus there is a curvilinear relationship with influence such that low influence and super high influence tweets are posted by regular users; whereas highly influential tweets are posted by news agencies and government officials.

\subsection*{What types of users cite ``fake news'' sites or discuss disinformation story-lines?}

To study the identity of users who are talking about ``fake news'' URLs, we first retrieve all the tweets that contained a URL to a news site (fake or real), then we apply our user profiling system to these users. In total, there are 3085 source tweets containing black news URLs, 53531 source tweets containing red news URLs, and 61179 source tweets containing orange news URLs, and 796267 containing real news URLs. The number of retweets containing such URLs are 32305 (black), 151286 (red), 205409 (orange), and 2739257 (real).

In Table \ref{ch6_news_identity}, we listed the number of tweets by each type of user that contain a URL to a news site by level of credibility. As shown in this table,  news sites with different levels of credibility attract different types of users  ($\chi^2=8832.1, p<0.001, df=18$). About 90\% of the tweets containing these ``fake news'' URLs are initiated by regular users. Most of those tweets contain links to red or orange sites. In contrast, the real news sites are linked to by governments and individual news reporters. 

\begin{figure*}[!h]
    \centering
    \includegraphics[width=0.75\textwidth]{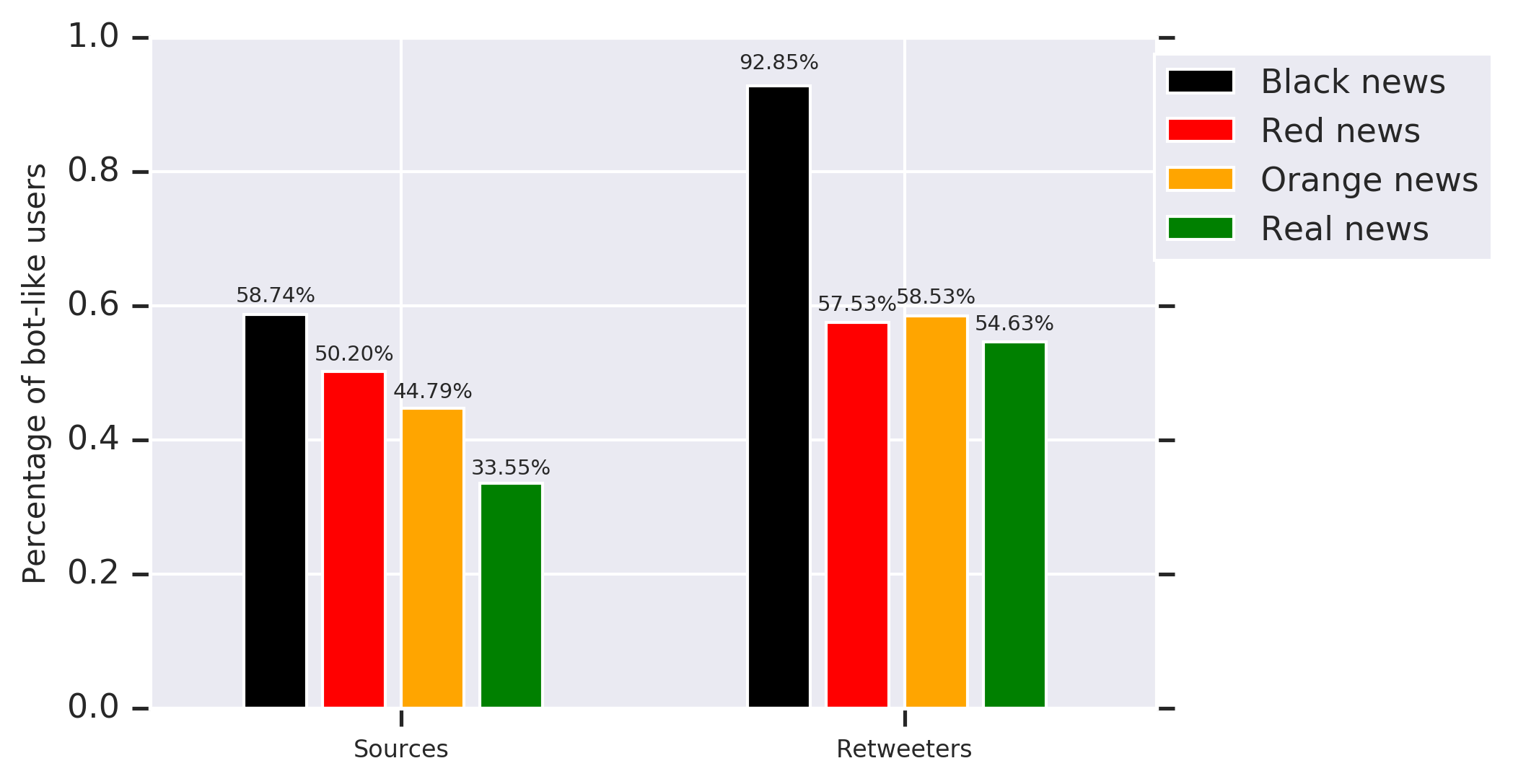}
    \includegraphics[width=0.75\textwidth]{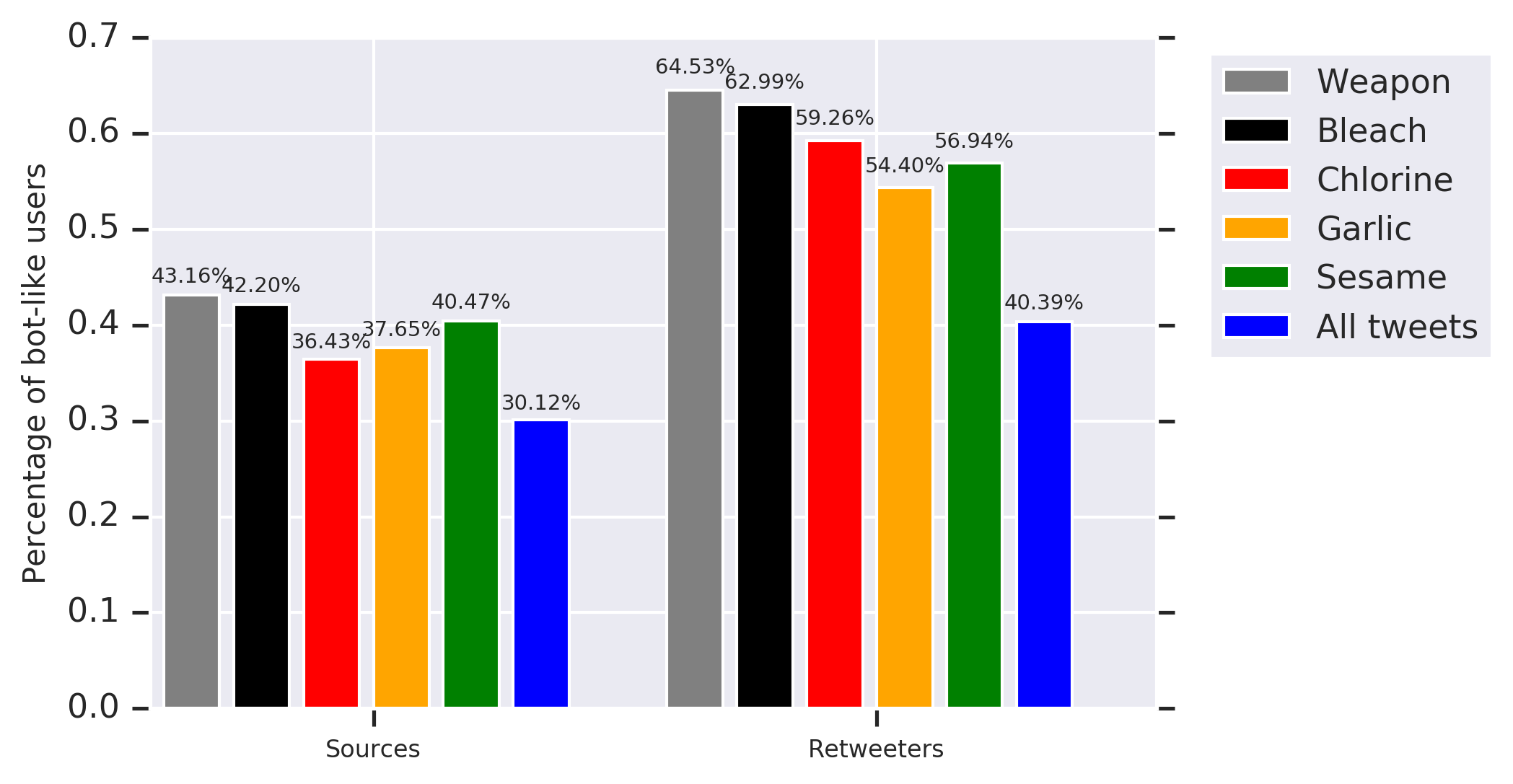}
    \caption{Percentage of bot-like source users and retweeters who share news URLs (top). Percentage of bot-like source users and retweeters who talked about misinformation (bottom). }
    \label{ch6_bot}
\end{figure*}

We also show the percentage of tweets sent by each type of user with bot-like behavior by levels of credibility in Table \ref{ch6_news_identity}. We find that the lower the credibility of the news site being linked to, the more likely the sender of the tweet is a bot. For example, 58.74\% of the source users sending a URL for a black news site appear to be bots as shown in Fig. \ref{ch6_bot}. A majority of retweeters of black sites are predicted as bot accounts, which indicates that a large group of automatically operated accounts are trying to promote these news. We also note that some of the news agencies, government and news reporter users appear to be bots. None of the accounts labeled as possible bots are verified accounts.  There are several possible reasons for this: a) there are news bots and propaganda bots that are employed by some news agencies and various official account, b) an account where multiple people send out the tweets can appear as a bot, and c) despite 95.7\% precision the bot-hunter program may be making errors. Nonetheless, the results suggest that there may be a set of bots that were established precisely to spread information from the less credible sites - particularly the black news sites.

\begin{table}[!h]
\caption{Number of tweets by each type of user that contain a URL to a news site by levels of credibility. Number in parentheses represents the percentage of tweets by each type of user that also have bot-like behavior, e.g. among 2090 tweets citing black news by regular users, 71.15\% of them are posted by bot-like accounts. }
\label{ch6_news_identity}
\centering
\resizebox{\columnwidth}{!}{%

\begin{tabular}{|l|l|l|l|l|}
\hline
            & Regular           & News agency      & Government       & News reporter    \\ \hline
Black news  & 2,090 (71.15\%)   & 409 (88.26\%)    & 36 (80.56\%)     & 27 (77.78\%)     \\ \hline
Red news    & 45,200 (66.74\%)  & 3,155 (85.93\%)  & 490 (60.00\%)    & 357 (63.03\%)    \\ \hline
Orange news & 45,503 (55.59\%)  & 4,260 (56.15\%)  & 821 (71.13\%)    & 634 (46.53\%)    \\ \hline
Real news   & 509,188 (47.62\%) & 68,044 (61.19\%) & 15,430 (36.42\%) & 45,547 (23.96\%) \\ \hline
\end{tabular}
}
\end{table}

What types of users are discussing the disinformation story-lines varies by story-line ($\chi^2=5233.8, p<0.001, df=30$) as can be seen in Table \ref{ch6_misinfo_identity}. The bio-weapon conspiracy is the most widely spread story-line as 34,301 unique users tweet about this story. Bleach is the most popular story-line concerning a false cure. A manual examination of the content of bleach tweets, showed that many of them appeared to be jokes or satirical responses to the original disinformation story. Fewer regular users spread the remaining story-lines.  Many news agencies and government officials such as WHO are sending tweets trying to refute the original disinformation story-line. 

\begin{table}[]
\caption{Number of tweets by each type of user that mention one of the story-lines. Number if parentheses represents percentage of tweets by each type of user with bot-like behavior that contain one of the story-lines, e.g. among 45,791 tweets mentioning bio-weapon by regular users, 47.07\% of them are posted by bot-like accounts.}
\label{ch6_misinfo_identity}
\resizebox{\columnwidth}{!}{%
\begin{tabular}{|l|l|l|l|l|}
\hline
Misinformation   & Regular             & News agency         & Government        & News reporter     \\ \hline
Bio-weapon       & 45,791 (47.07\%)    & 1,956 (68.71\%)     & 842 (70.07\%)     & 1,192 (71.47\%)   \\ \hline
Bleach           & 5,826 (42.76\%)     & 280 (47.86\%)       & 72 (36.11\%)      & 142 (21.83\%)     \\ \hline
Chlorine dioxide & 262 (35.11\%)       & 58 (22.41\%)        & 4 (50.00\%)       & 5 (60.00\%)       \\ \hline
Garlic           & 2,316 (40.85\%)     & 385 (56.62\%)       & 79 (29.11\%)      & 100 (28.00\%)     \\ \hline
Sesame           & 279 (38.71\%)       & 33 (66.67\%)        & 27 (22.22\%)      & 22 (18.18\%)      \\ \hline
All tweets       & 6,516,340 (42.35\%) & 1,102,842 (59.75\%) & 185,731 (39.01\%) & 263,939 (26.55\%) \\ \hline
\end{tabular}
}
\end{table}

In many cases, though, it appears that it is bots sending tweets regarding these disinformation story-lines.  In Table \ref{ch6_misinfo_identity} we see that there is not a simple pattern to the bot activity.  We do find that there are more bot-like users spreading the bio-weapon story-line. This suggest that a set of bots may have been established to mimic authoritative sites to spread this information.

\begin{figure*}[!h]
    \centering
    \includegraphics[width=0.75\textwidth]{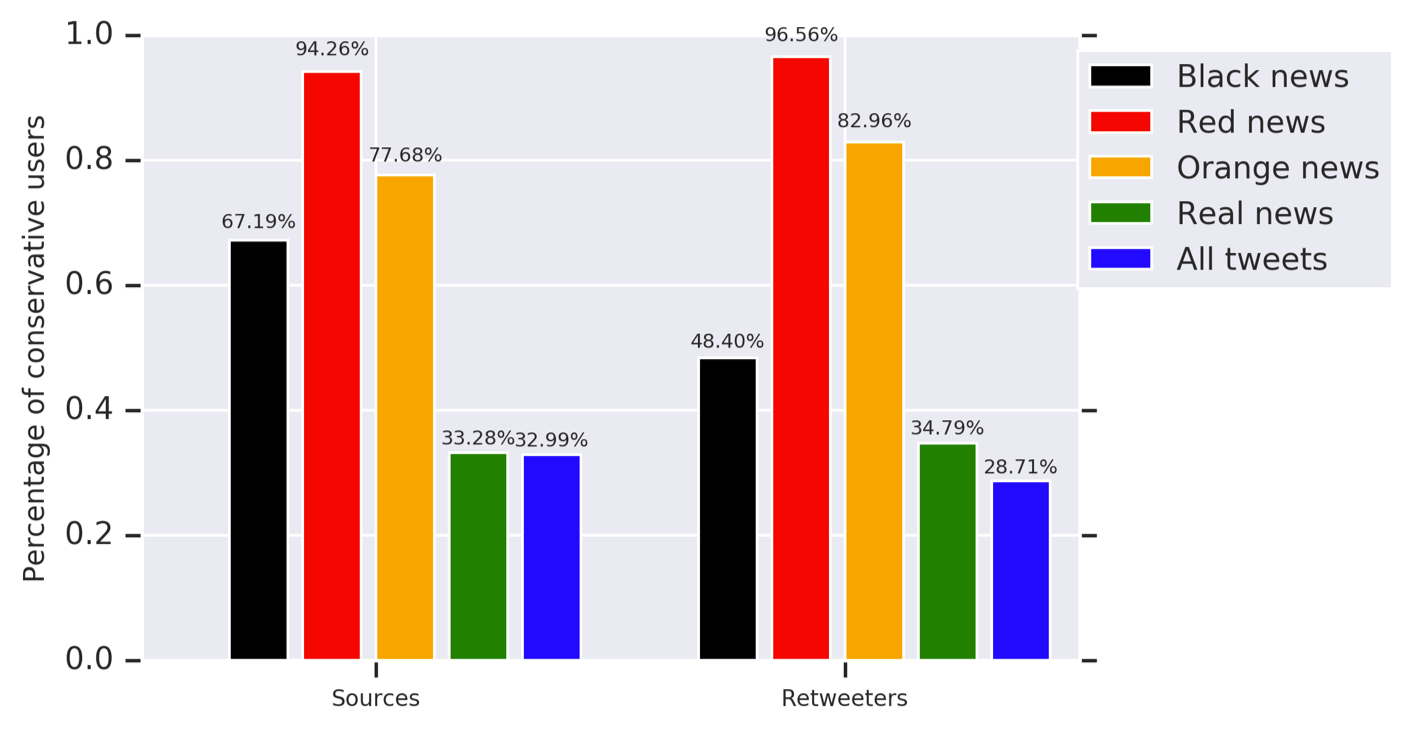}
    \includegraphics[width=0.75\textwidth]{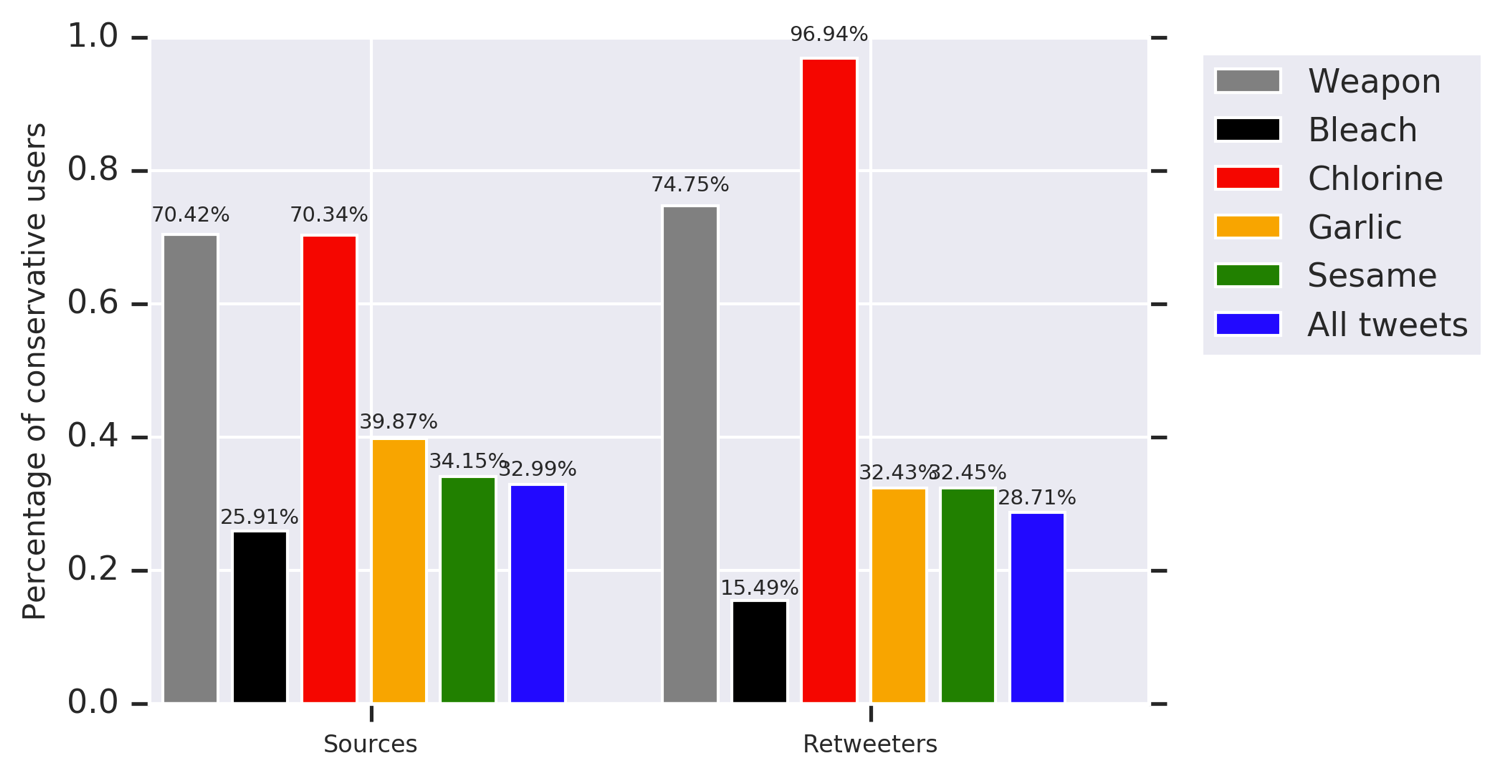}
    \caption{Percentage of conservative source users and retweeters who share news URLs (top). Percentage of conservative source users and retweeters who talked about misinformation (bottom). }
    \label{ch6_politic}
\end{figure*}

For the political orientation, Fig. \ref{ch6_politic} shows most people who tweet and retweet ``fake news'' URL are more leaning towards conservative users. 33.28\% of people tweeting real news URLs are labeled as conservative users, while 82.45\% of people sharing ``fake news'' URLs are labeled conservative users. For each misinformation story, the percentage varies case by case. The widely spread bio-weapon story tends to attract a higher percentage of conservative users than liberal users. For the other cure stories, most of users discussing them are labeled as liberal users. Interestingly, even though chlorine dioxide is one kind of bleach, people who tweet about chlorine dioxide differ significantly from the bleach case, and the tweets are less likely to be jokes.

\subsection*{Where in the world are those who discuss low credibility information?}
To measure users' geographical distribution, we combine geotags in user timelines with our country-level location predictions. For each user collected with timeline data, we look at the geotag in each tweet. Among 11,951,739 users, there are 791,830 individuals with geotags in their timelines. The home country of these geotagged users is determined by a majority vote of these geotags. For the remaining users, we apply our country-level location predictor to get their home countries. 

Because these news lists are mainly compiled by English speakers, we only measured the country distributions among English speaking users to avoid potential language bias. As shown in Table \ref{ch6_country_fake_source}, United States, United Kingdom, and Canada are the three countries from which most tweets with ``fake news'' URLs originate.  Because of the language bias, there are many more tweets mentioning ``fake news''  and misinformation stories from countries mainly speaking in English such as United States and United Kingdom.  73.26\% of the English speakers who posted tweets with ``fake news''  URLs came from United States. The percentage is also much higher than the U.S. users who posted real news URLs. To partially control for this bias, we normalize the number of users sharing ``fake news''  URLs by the total number of users in each country. The last column of Table \ref{ch6_country_fake_source} shows the normalized results. The probability for each U.S. user posting a ``fake news''  URL is 2.28\%, while the probabilities for Philippines and Malaysia are 0.13\% and 0.17\%. Even though many users in United Kingdom shared ``fake news''  URLs, the normalized number is much lower than countries like the United States and Canada. Users mentioning ``fake news''  URLs are more different from the underlying population than users mentioning real news. We use KL-divergence to measure the distance from country distribution of source users posting ``fake news''  URLs to the country distribution of all English sources. The distance is 0.144 in this case. On the contrary, the KL-divergence from country distribution of sources posting real news URLs to the country distribution of English sources is only 0.075. 

In Table \ref{ch6_country_fake_retweeter}, we also show the country distribution for users who retweet these ``fake news''  URLs. Again, a majority of retweeters are from the United States, followed by the United Kingdom and Canada. After we normalize the number of retweeters of ``fake news''  URLs by the total number of retweeters in each country, we can see that 3.79\% of users in United States have at least retweeted one tweet with ``fake news''  URLs, which is much higher than the average percentage. Even though the United Kingdom has the second most English speaking retweeters, the probability for users in the United Kingdom retweeting ``fake news''  URLs is still lower than the average probability.
Again, users retweeting ``fake news''  URLs are more different from the underlying population than are users mentioning real news.
The KL-divergence from country distribution of retweeters of ``fake news''  URLs to English retweeters is 0.192, while the distance from real news retweeters to English retweeters is only 0.096. 

\begin{table}[]
\caption{Country distribution of English speaking users who posted tweets with news URLs. We only show top 10 countries with the most English source users in this table.}
\label{ch6_country_fake_source}
\resizebox{\columnwidth}{!}{%
\begin{tabular}{|l|l|l|l|l|}
\hline
Country/region & \begin{tabular}[c]{@{}l@{}}\# of EN users posting\\ ``fake news''  URLs \end{tabular} & \begin{tabular}[c]{@{}l@{}}\# of EN users posting\\ real news URLs\end{tabular} & \# of EN source users & \begin{tabular}[c]{@{}l@{}}\% of EN users posting\\ ``fake news''  URLs per country\end{tabular} \\ \hline
United States  & 24,552 (73.26\%) & 146,875 (64.11\%) & 1,077,431 (51.55\%) & 2.28\% \\ \hline
United Kingdom & 2,599 (7.75\%)   & 23,988 (10.47\%)  & 255,779 (12.24\%)   & 1.02\% \\ \hline
Philippines    & 112 (0.33\%)     & 2,483 (1.08\%)    & 89,480 (4.28\%)     & 0.13\% \\ \hline
India          & 334 (1.00\%)     & 4,674 (2.04\%)    & 83,030 (3.97\%)     & 0.40\% \\ \hline
Canada         & 1,368 (4.08\%)   & 10,228 (4.46\%)   & 77,344 (3.70\%)     & 1.77\% \\ \hline
Nigeria        & 677 (2.02\%)     & 1,263 (0.55\%)    & 59,862 (2.86\%)     & 1.13\% \\ \hline
Australia      & 607 (1.81\%)     & 6,755 (2.95\%)    & 52,526 (2.51\%)     & 1.16\% \\ \hline
South Africa   & 339 (1.01\%)     & 1,261 (0.55\%)    & 26,308 (1.26\%)     & 1.29\% \\ \hline
Malaysia       & 39 (0.12\%)      & 961 (0.42\%)      & 22,869 (1.09\%)     & 0.17\% \\ \hline
Kenya          & 83 (0.25\%)      & 900 (0.39\%)      & 19,486 (0.93\%)     & 0.43\% \\ \hline
total          & 33,514           & 229,111           & 2,089,892           & 1.60\% \\ \hline
\end{tabular}
}
\end{table}

\begin{table}[]
\caption{Country distribution of English speaking users who retweeted tweets with news URLs. We only show top 10 countries with the most English retweeters in this table.}
\label{ch6_country_fake_retweeter}
\resizebox{\columnwidth}{!}{%
\begin{tabular}{|l|l|l|l|l|}
\hline
Country/region & \begin{tabular}[c]{@{}l@{}}\# of EN retweeters \\of ``fake news''  URLs \end{tabular} & \begin{tabular}[c]{@{}l@{}}\# of EN retweeters\\ of real news URLs\end{tabular} & \# of EN retweeters & \begin{tabular}[c]{@{}l@{}}\% of EN retweeters of \\fake news URLs per country\end{tabular} \\ \hline
United States  & 79,157 (75.52\%) & 343,418 (64.18\%) & 2,088,187 (49.88\%) & 3.79\% \\ \hline
United Kingdom & 6,667 (6.36\%)   & 49,222 (9.20\%)   & 441,331 (10.54\%)   & 1.51\% \\ \hline
India          & 1,642 (1.57\%)   & 13,277 (2.48\%)   & 206,542 (4.93\%)    & 0.79\% \\ \hline
Philippines    & 241 (0.23\%)     & 11,960 (2.24\%)   & 204,856 (4.89\%)    & 0.12\% \\ \hline
Nigeria        & 2,953 (2.82\%)   & 5,108 (0.95\%)    & 135,306 (3.23\%)    & 2.18\% \\ \hline
Canada         & 3,586 (3.42\%)   & 21,614 (4.04\%)   & 135,083 (3.23\%)    & 2.65\% \\ \hline
Malaysia       & 380 (0.36\%)     & 5,973 (1.12\%)    & 118,361 (2.83\%)    & 0.32\% \\ \hline
Australia      & 1,393 (1.33\%)   & 12,200 (2.28\%)   & 69,699 (1.66\%)     & 2.00\% \\ \hline
Indonesia      & 98 (0.09\%)      & 4,781 (0.89\%)    & 62,147 (1.48\%)     & 0.16\% \\ \hline
South Africa   & 1,093 (1.04\%)   & 2,875 (0.54\%)    & 57,511 (1.37\%)     & 1.90\% \\ \hline
total          & 104,811          & 535,113           & 4,186,548           & 2.50\% \\ \hline
\end{tabular}
}
\end{table}

We have similar observation on the country distribution of misinformation, as shown in Table \ref{ch6_country_misinfo}. Most of users involved in the conversation about misinformation are from United States, followed by United Kingdom, Canada. Among these top 10 countries, people from United States, Canada are more likely to tweet or retweet misinformation. Among 1,077,431 U.S. users who posted tweets, 2.16\% of them posted tweets mentioning misinformation stories. And 4.49\% of retweeters in U.S. have retweeted tweets talking misinformation stories. One thing we want to note is that even though only 11,734 and 32,643 English-speaking tweeters and retweeters are from Hong Kong, 2.86\% tweeters and 6.99\% retweeters from Hong Kong have tweeted or retweeted tweets mentioning misinformation phrases.

\begin{table}[]
\caption{Country distribution of English speaking users who are involved in the conversation of misinformation. We only show top 10 countries with the most English source users in this table.}
\label{ch6_country_misinfo}
\resizebox{\columnwidth}{!}{%
\begin{tabular}{|l|l|l|l|l|l|l|}
\hline
Country/Region & \begin{tabular}[c]{@{}l@{}}\# of EN sources\\talking misinfo.\end{tabular} & \# of EN sources  & \begin{tabular}[c]{@{}l@{}}\% of EN sources\\talking misinfo. \\ per country\end{tabular} & \begin{tabular}[c]{@{}l@{}}\# of EN retweeters\\ of misinfo. \end{tabular} & \begin{tabular}[c]{@{}l@{}}\# of EN\\ retweeters\end{tabular} & \begin{tabular}[c]{@{}l@{}}\% of retweeters\\of misinfo.  \\ per country\end{tabular} \\ \hline
United States  & 23,234 (63.40\%) & 1,077,431 (51.55\%) & 2.16\% & 93,696 (67.54\%) & 2,088,187 (49.88\%) & 4.49\% \\ \hline
United Kingdom & 2,327 (6.35\%)   & 255,779 (12.24\%)   & 0.91\% & 7,484 (5.39\%)   & 441,331 (10.54\%)   & 1.70\% \\ \hline
Philippines    & 1,142 (3.12\%)   & 89,480 (4.28\%)     & 1.28\% & 2,401 (1.73\%)   & 204,856 (4.89\%)    & 1.17\% \\ \hline
India          & 1,351 (3.69\%)   & 83,030 (3.97\%)     & 1.63\% & 5,330 (3.84\%)   & 206,542 (4.93\%)    & 2.58\% \\ \hline
Canada         & 1,541 (4.21\%)   & 77,344 (3.70\%)     & 1.99\% & 4,791 (3.45\%)   & 135,083 (3.23\%)    & 3.55\% \\ \hline
Nigeria        & 1,080 (2.95\%)   & 59,862 (2.86\%)     & 1.80\% & 4,181 (3.01\%)   & 135,306 (3.23\%)    & 3.09\% \\ \hline
Australia      & 879 (2.40\%)     & 52,526 (2.51\%)     & 1.67\% & 1,962 (1.41\%)   & 69,699 (1.66\%)     & 2.81\% \\ \hline
South Africa   & 359 (0.98\%)     & 26,308 (1.26\%)     & 1.36\% & 1,539 (1.11\%)   & 57,511 (1.37\%)     & 2.68\% \\ \hline
Malaysia       & 219 (0.60\%)     & 22,869 (1.09\%)     & 0.96\% & 2,009 (1.45\%)   & 118,361 (2.83\%)    & 1.70\% \\ \hline
Kenya          & 341 (0.93\%)     & 19,486 (0.93\%)     & 1.75\% & 1,773 (1.28\%)   & 39,558 (0.94\%)     & 4.48\% \\ \hline
total          & 36,645           & 2,089,892           & 1.75\% & 138,723          & 4,186,548           & 3.31\% \\ \hline
\end{tabular}
}
\end{table}

\begin{figure*}[!h]
    \centering
    \includegraphics[width=0.75\textwidth]{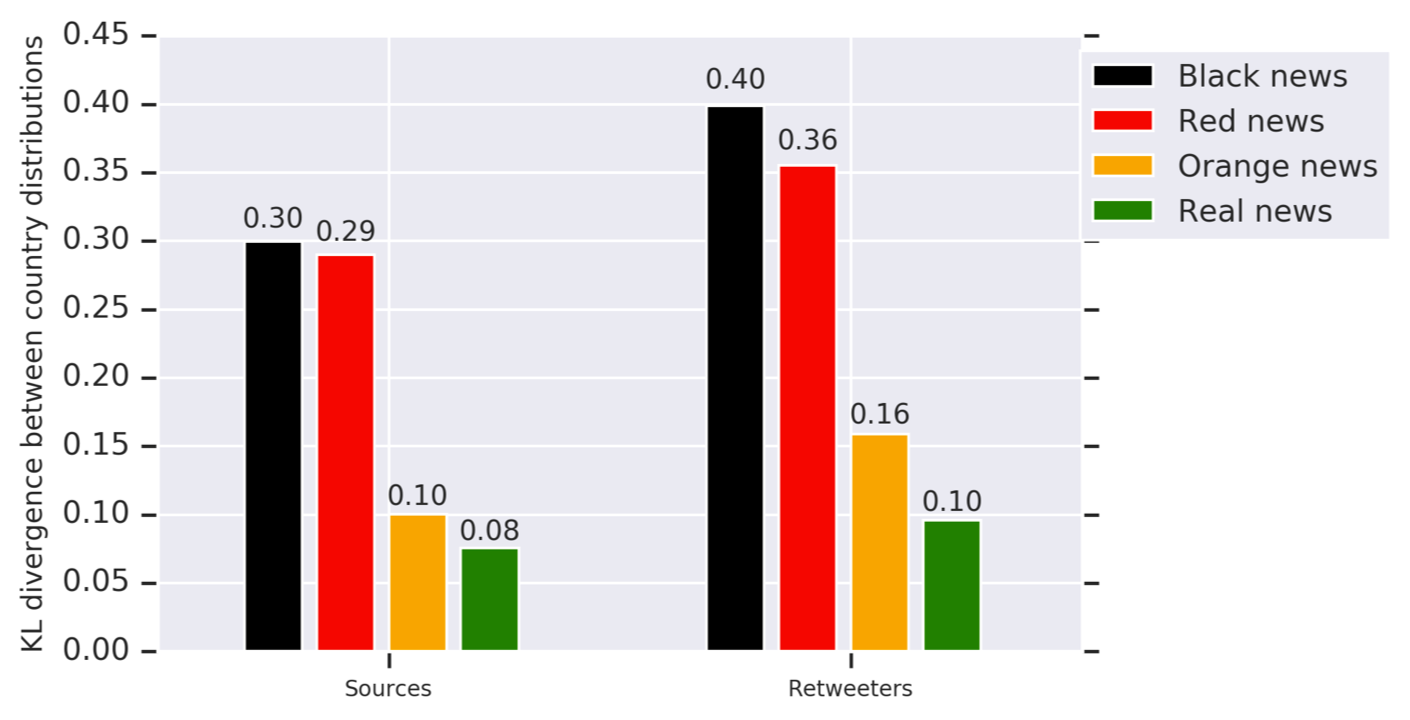}
    \includegraphics[width=0.75\textwidth]{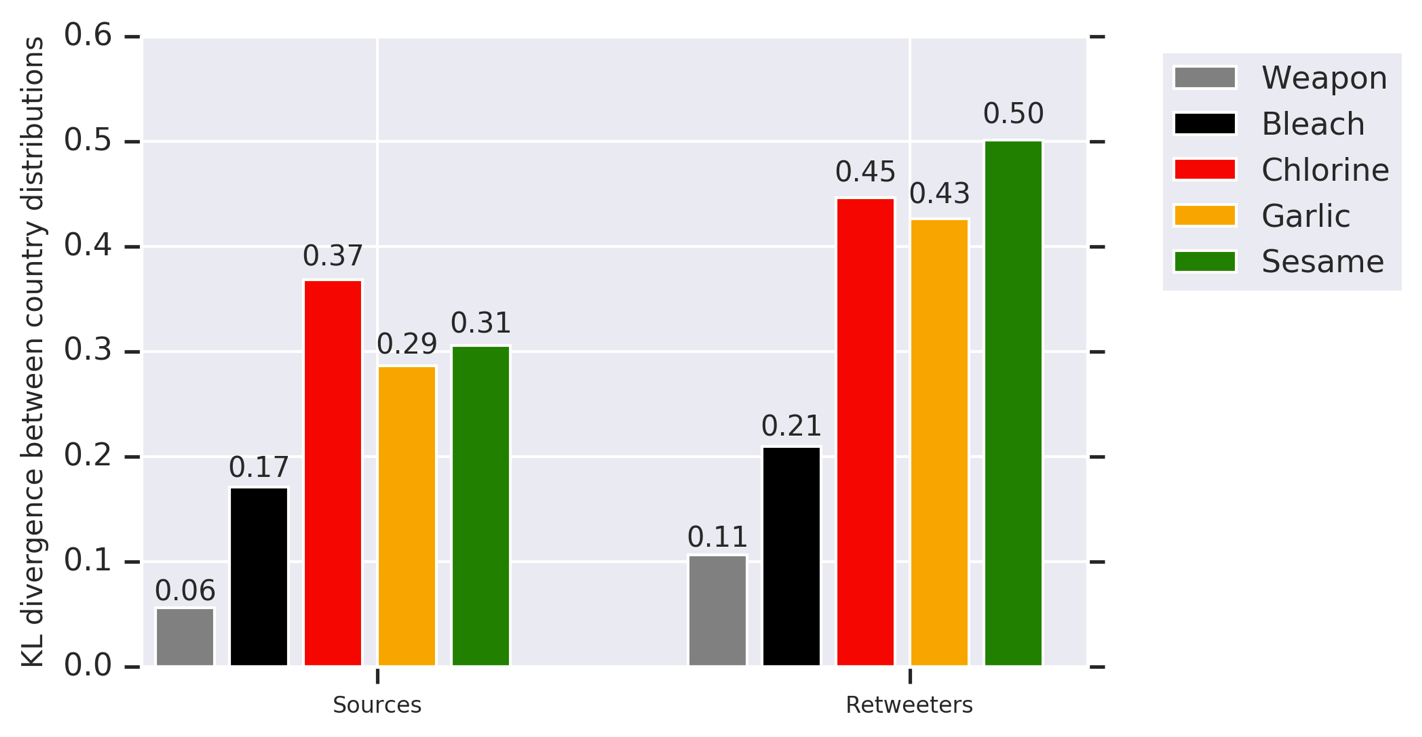}
    \caption{KL-divergence between country distributions of EN users sharing news and all EN users (above). KL-divergence between country distributions of EN users talking misinformation stories and all EN users (bottom). }
    \label{ch6_kl}
\end{figure*}

We plot the Kl-divergence between the country distribution of English speaking users who share specific news and the country distribution of all English speakers in Fig. \ref{ch6_kl}. As shown in this figure, the population of who sharing real news is the closest one to the underlying one. The less credible the news sites, the bigger of the population difference. In the bottom of Fig. \ref{ch6_kl}, we also show the same plot for all misinformation stories. The country distribution of source users who talked about the bio-weapon conspiracy are very close to the one of underlying population. Among the cure stories, the bleach misinformation is the one spread most closest to the underlying population.

\subsection*{What is the global network for discussing low credibility information?}
\begin{figure*}[!h]
    \centering
    \includegraphics[width=0.75\textwidth]{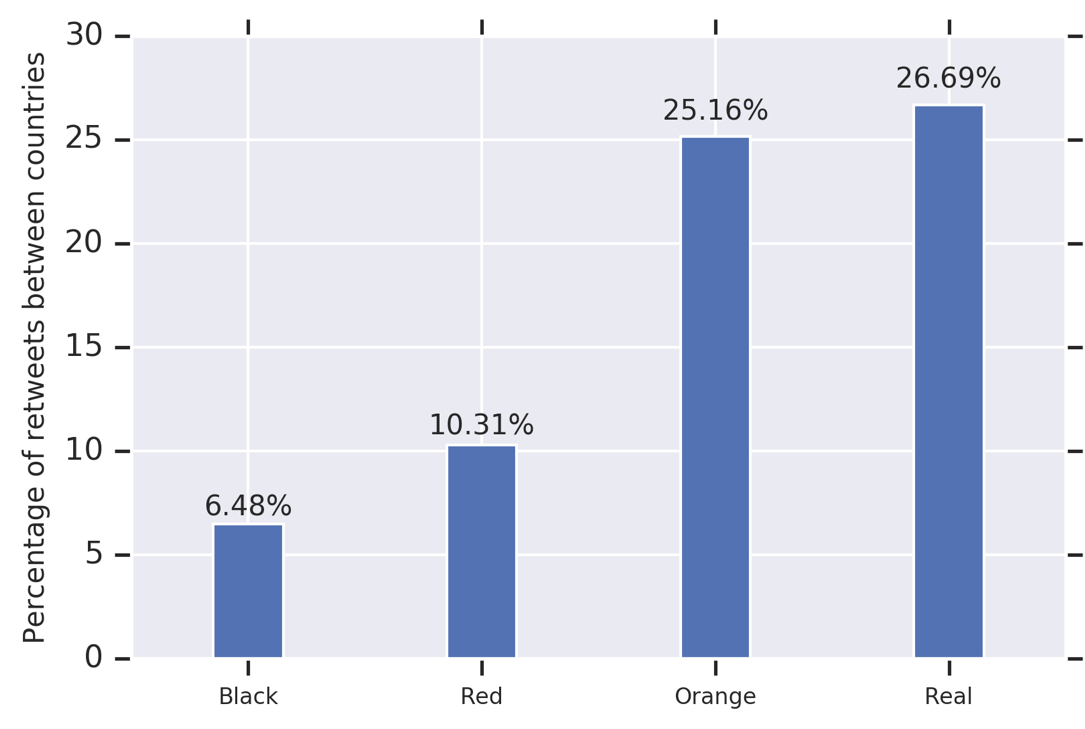}
    \includegraphics[width=0.75\textwidth]{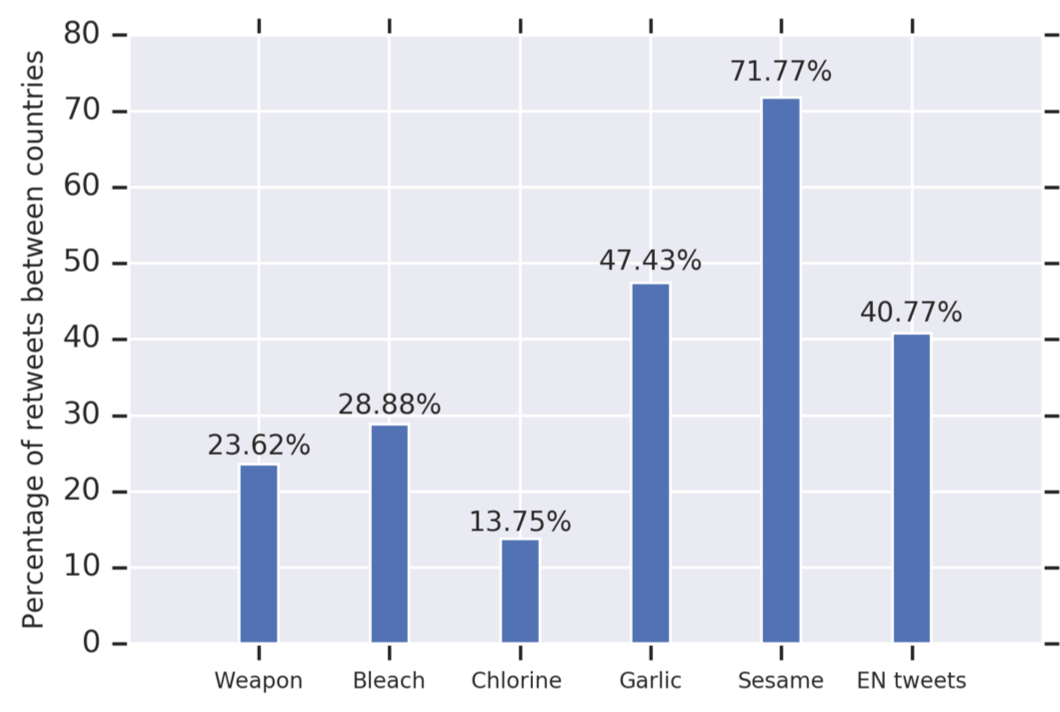}
    \caption{Percentages of retweets between countries for ``fake news''  sites and misinformation conversations.}
    \label{ch6_intra_country}
\end{figure*}

An important question that has received little attention is how information about low credibility websites and disinformation story-lines spread between countries. To examine this, we extracted the information flow among countries. If a user in country A retweets a tweet posted by a user in country B, then we add an edge from country B to country A. Figure 5 shows the percentage of retweets between countries for tweets with ``fake news''  URLs and misinformation conversations. As shown in the above Fig. \ref{ch6_intra_country}, 26.69\% of users who retweet tweets with real news URLs are from a different country than the source tweet. The percentage of inter-country retweets is much lower for the tweets containing URLs to less credible sites, especially for the black news sites. This demonstrates that tweets mentioning less credible news-sites tend to stay within the source country.  This helps explain why country distribution of users sharing these ``fake news'' URLs differ from the underlying population.

As for the misinformation stories, tweets talking about bio-weapon, bleach, chlorine dioxide are more likely to be retweeted by users from the same country. Tweets mentioning garlic and sesame are more likely to spread internationally. One reason for this is that global health agencies such as WHO posted several clarifications for these misinformation stories.

\begin{figure*}[!h]
    \centering
    \includegraphics[width=0.75\textwidth]{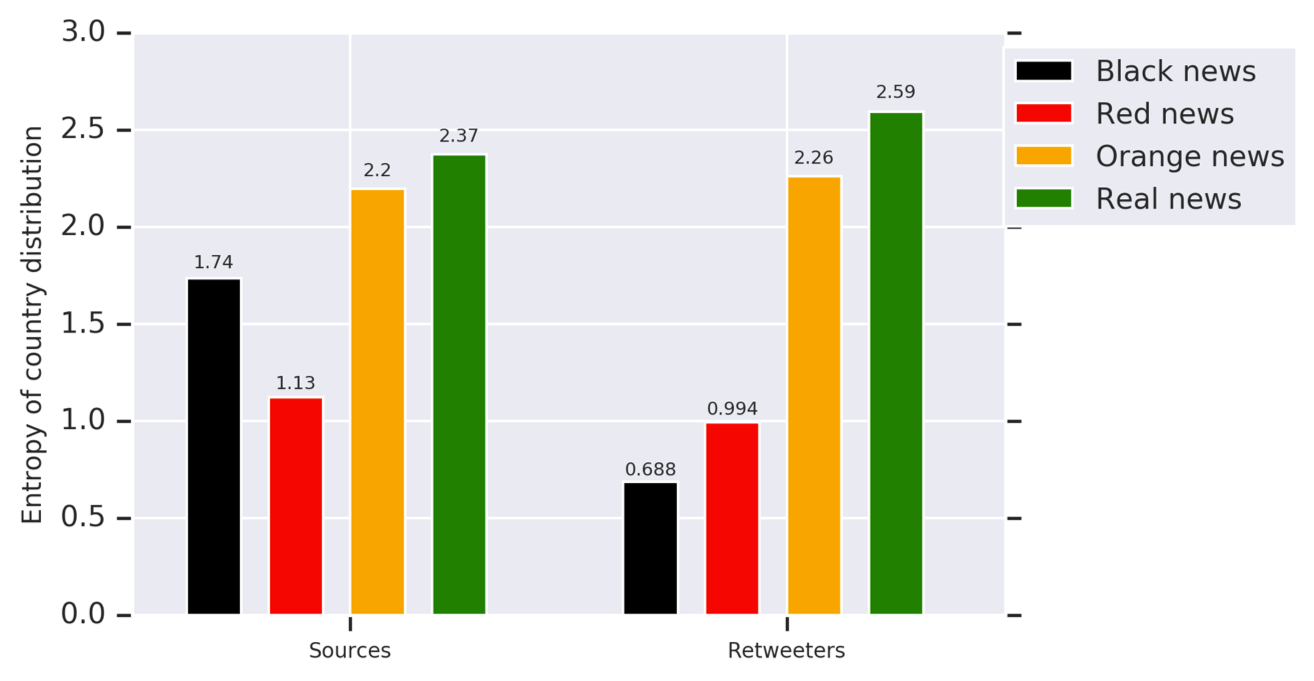}
    \includegraphics[width=0.75\textwidth]{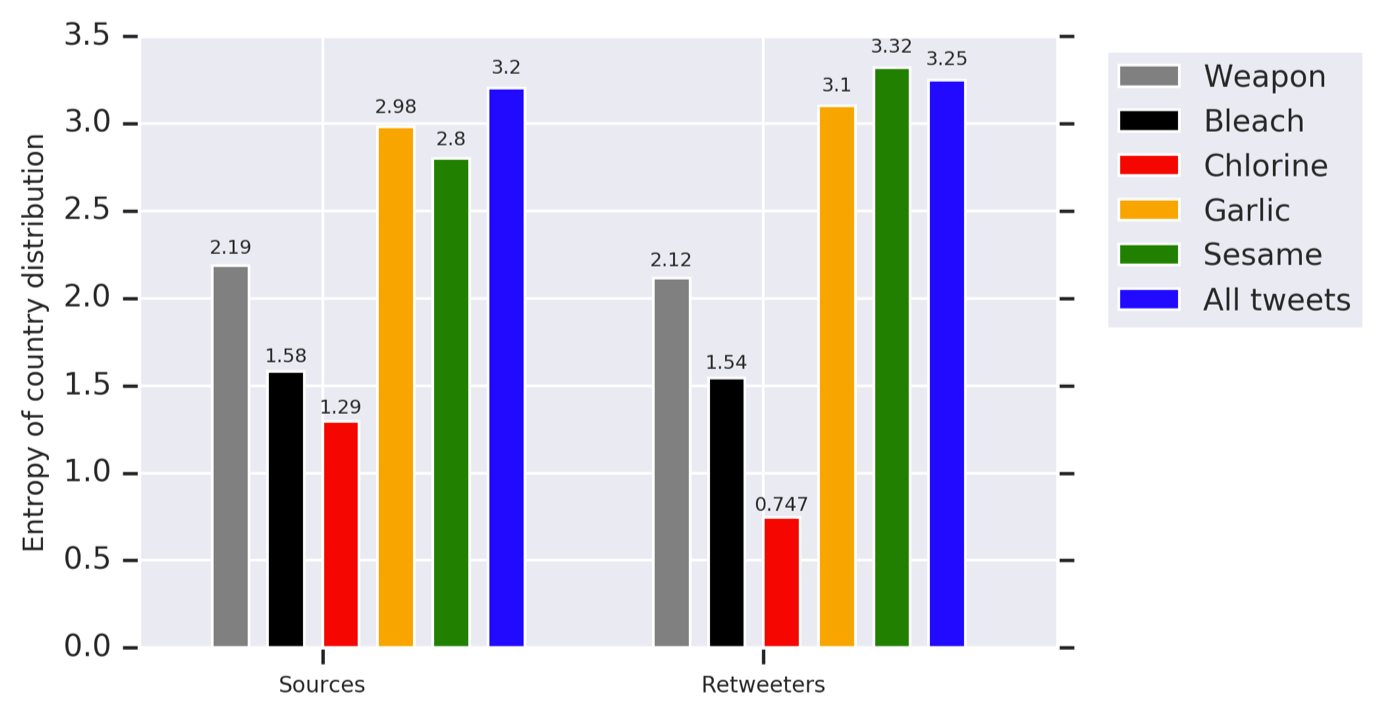}
    \caption{Entropy of country distribution for users who posted certain ``fake news''  URLs and misinformation stories.}
    \label{ch6_entropy}
\end{figure*}

As a result, conversations about real news and certain misinformation stories have high country diversity in their spread. Here, we use entropy of country distribution to measure the country diversity of Twitter users who shared each ``fake news''  sites and misinformation stories. As shown in Fig. \ref{ch6_entropy}, the country entropy of users who posted tweets with ``fake news''  URLs are much less than users who posted real news URLs. The black news and red news are constrained in the source countries. Similar effect also happens for misinformation stories. Tweets talking about garlic and sesame are spreading as diverse as normal tweets, while users talking about chlorine dioxide are highly concentrated in certain countries.

\begin{figure*}[!h]
    \centering
    \begin{subfigure}[b]{0.23\textwidth}
            \centering
            {\includegraphics[width=\textwidth]{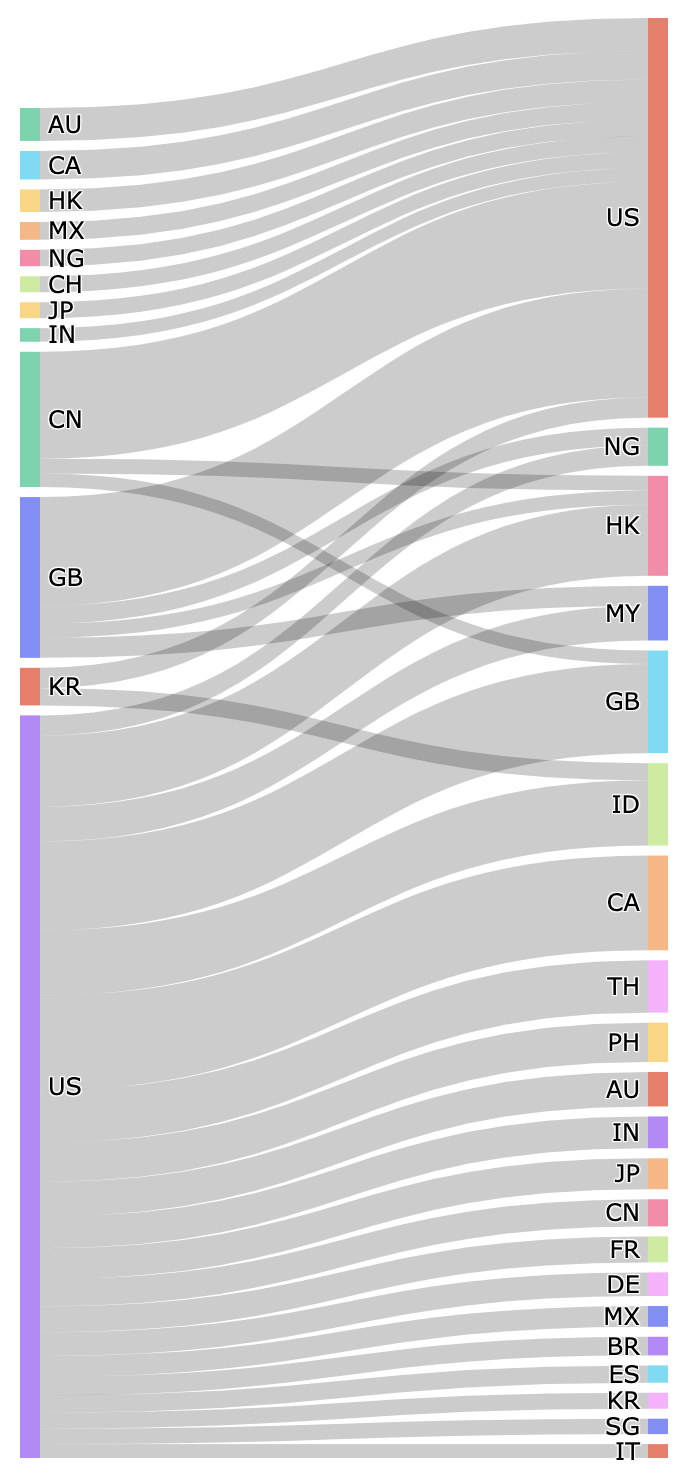}}
            \caption{Flow of EN tweets between countries}
             \label{flow:en}
    \end{subfigure}%
    \begin{subfigure}[b]{0.23\textwidth}
            \centering
            {\includegraphics[width=\textwidth]{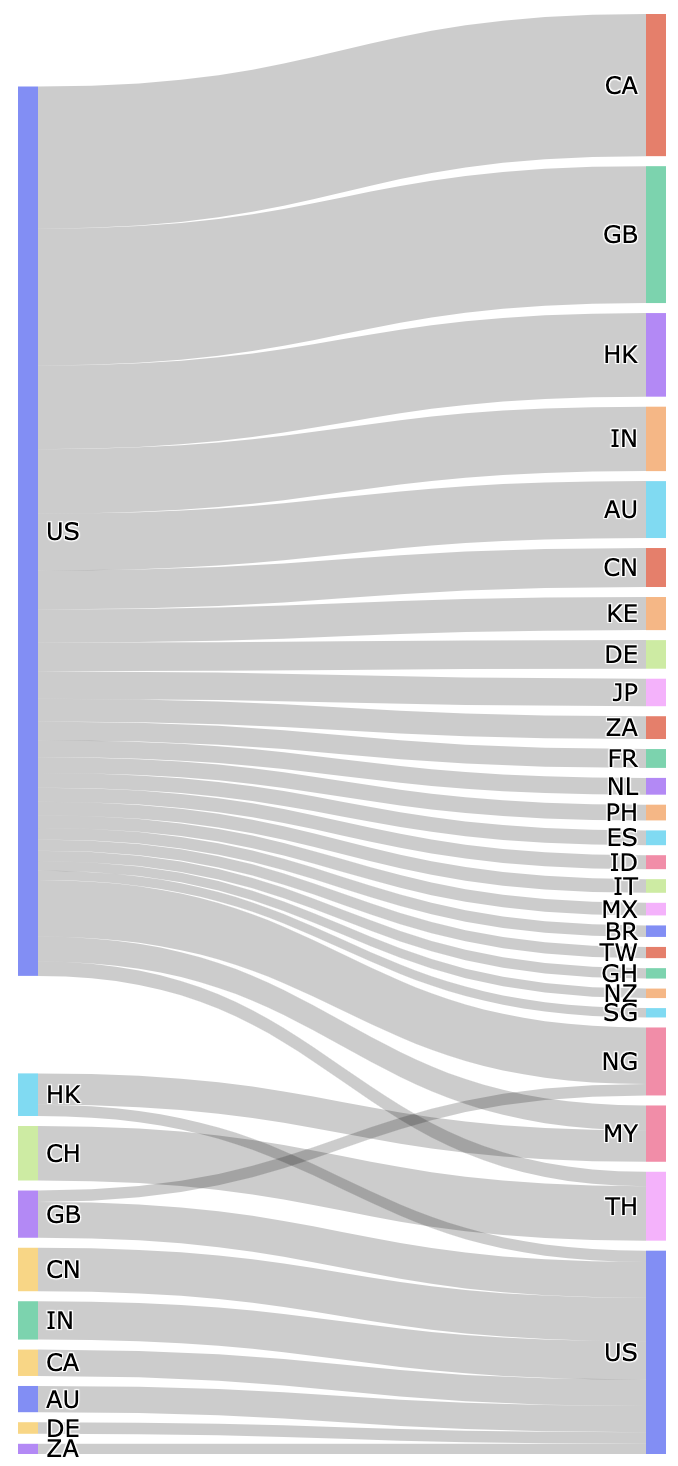}}
            \caption{Flow of tweets talking bioweapon}
             \label{flow:weapon}
    \end{subfigure}%
    \begin{subfigure}[b]{0.23\textwidth}
            \centering
            {\includegraphics[width=\textwidth]{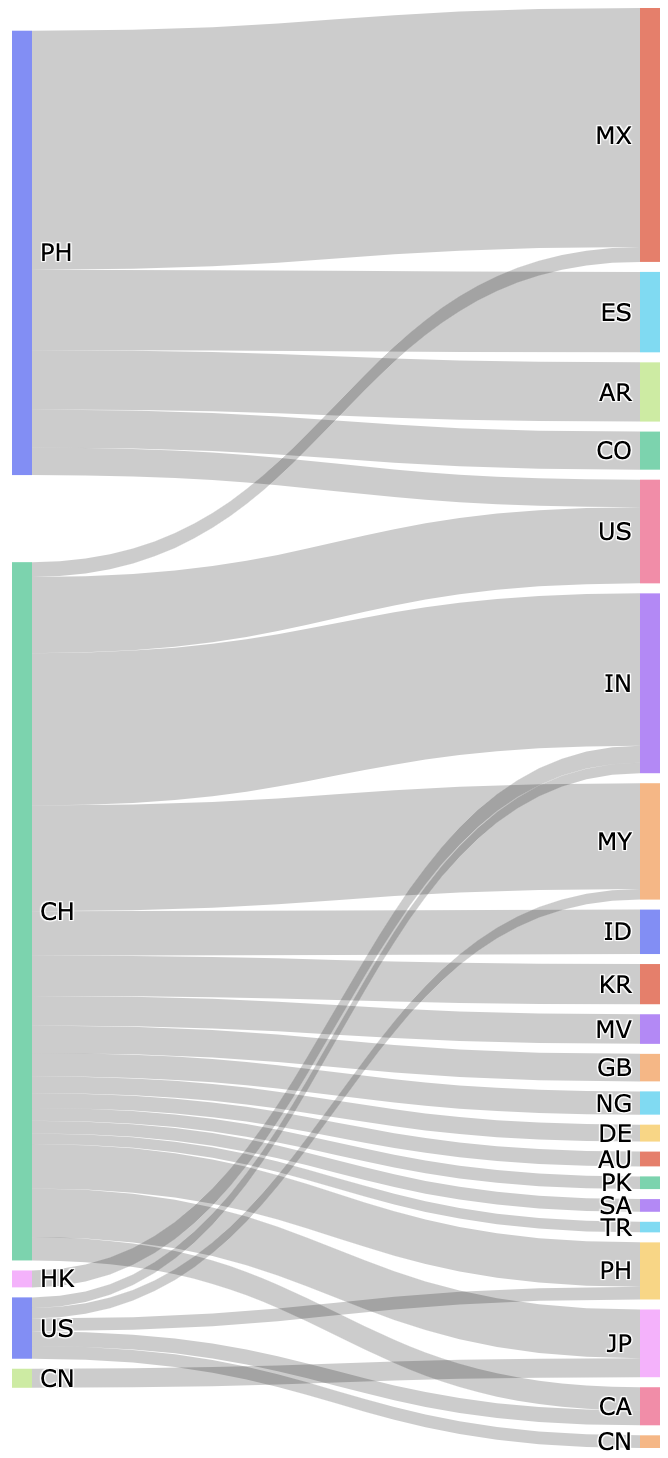}}
            \caption{Flow of tweets talking sesame}
             \label{flow:sesame}
    \end{subfigure}%
    \\
    \begin{subfigure}[b]{0.23\textwidth}
            \centering
            {\includegraphics[width=\textwidth]{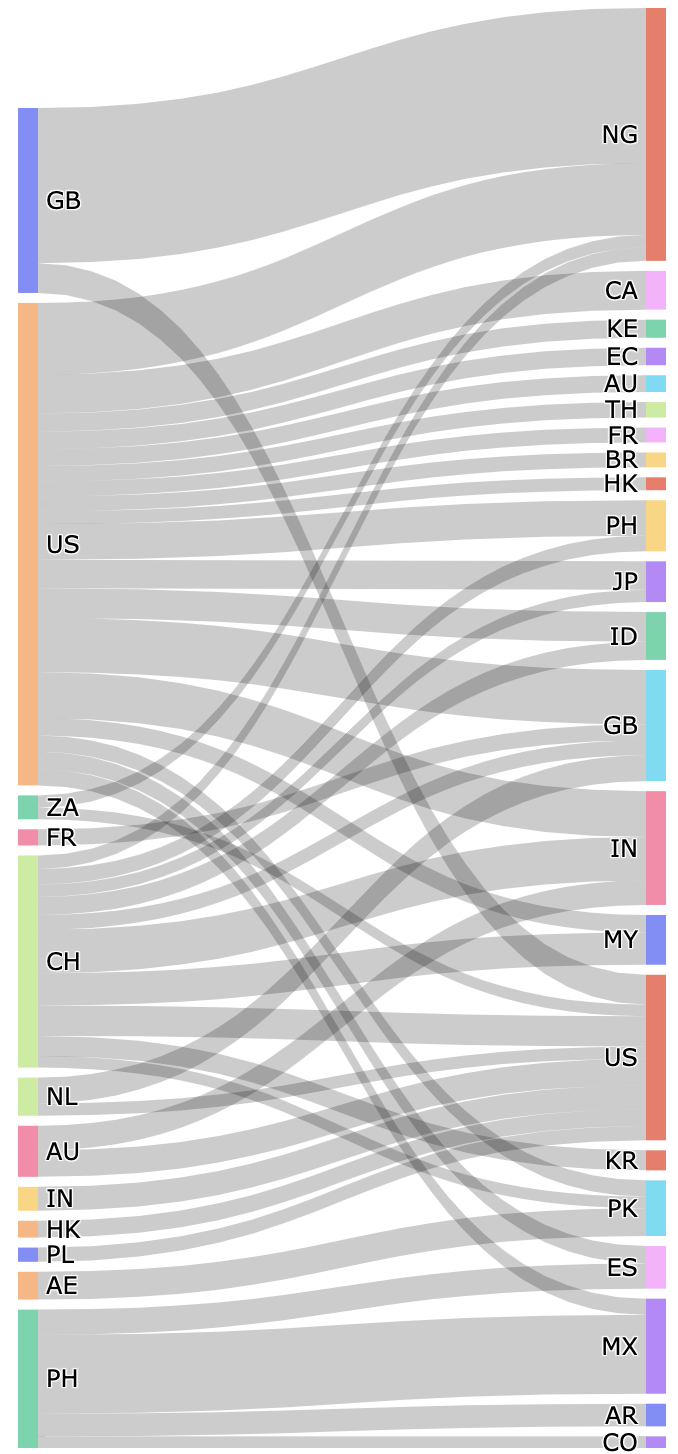}}
            \caption{Flow of tweets talking garlic}
             \label{flow:garlic}
    \end{subfigure}%
    \begin{subfigure}[b]{0.23\textwidth}
            \centering
            {\includegraphics[width=\textwidth]{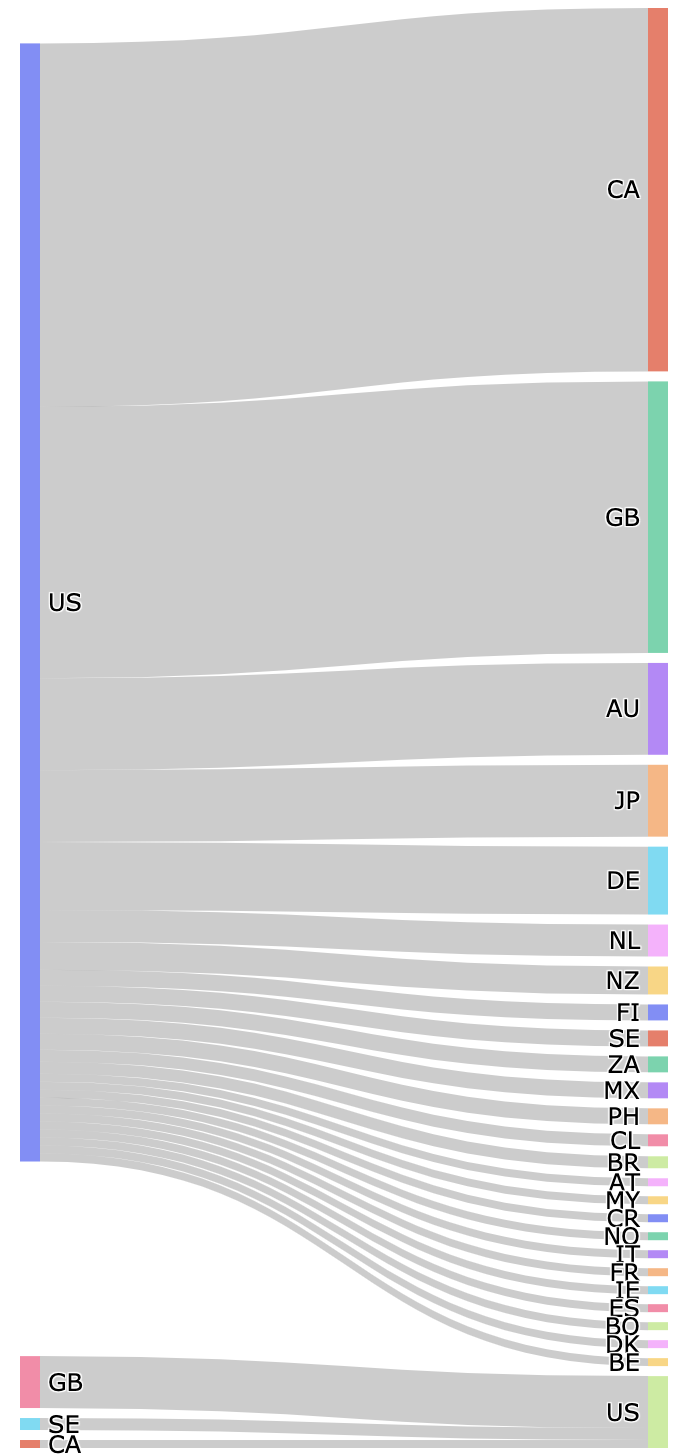}}
            \caption{Flow of tweets talking chlorine dioxide}
             \label{flow:chlorine}
    \end{subfigure}
    \begin{subfigure}[b]{0.23\textwidth}
            \centering
            {\includegraphics[width=\textwidth]{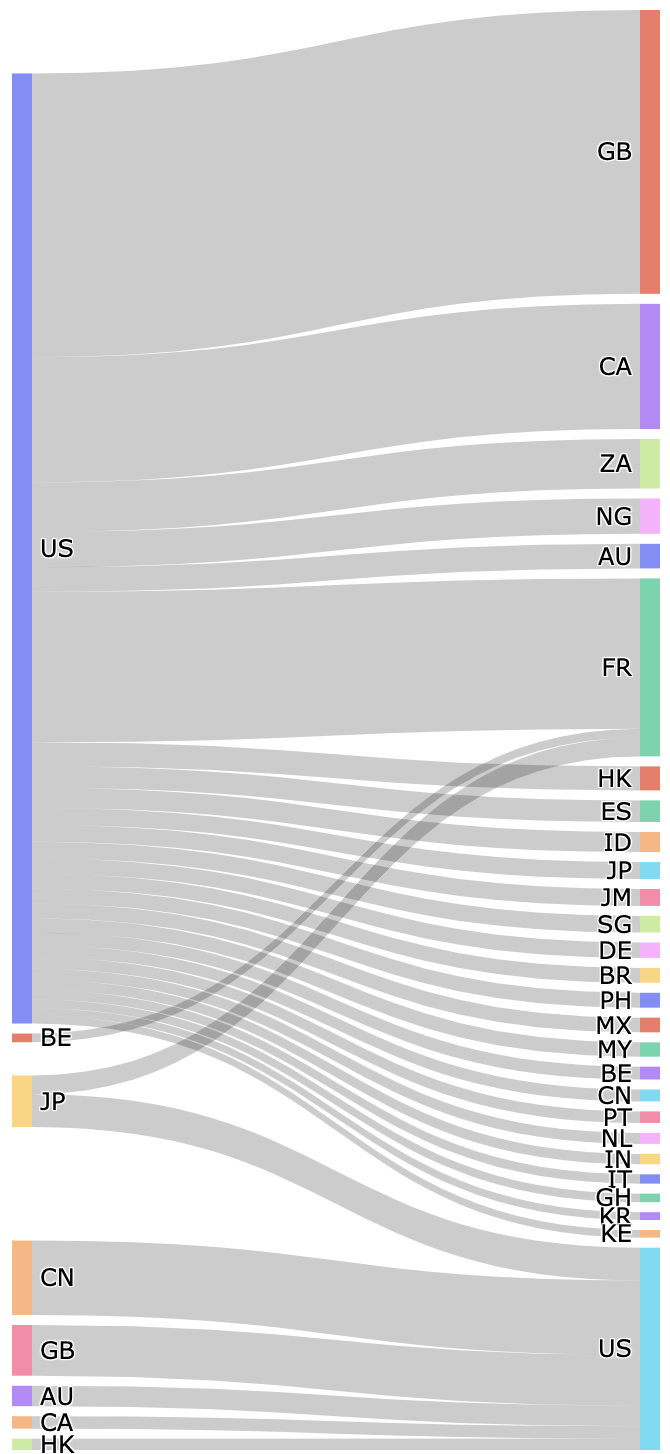}}
            \caption{Flow of tweets talking bleach}
             \label{flow:bleach}
    \end{subfigure}
    \caption{Information flows among countries. Retweets from the same country as the source are excluded in this figure. We use ISO 3166-1 alpha-2 country code.}
    \label{ch6_flow}
\end{figure*}

In Fig. \ref{ch6_flow}, we show the information flows among countries. The width of a flow is proportional to the percentage of retweets from country A to country B among retweets between all the countries. We only show information flows with percentages higher than 0.5\%. We also exclude the retweets inside the same country in this figure.  As shown in Fig. \ref{flow:en}, United States contributing the most of the retweeting flow between countries. Out of 11,092,477 inter-country retweets, 4,034,985 are from United States (36.38\%).
A great deal of information moves from the United States to other countries particularly the United Kingdom and Hong Kong. The situations become more extreme in the cases of bio-weapon (39,535 out of 69,389, 56.98\%), bleach (5,803 out of 8,144, 71.25\%), and chlorine dioxide (292 out of 326, 89.6\%). Switzerland, where WHO is located, plays a larger role in the flow of information about sesame and garlic. For example, for the sesame story-line there are 823 retweets between countries, 415 of them are from Switzerland (50.43\%). All of these retweets are people retweeting WHO.

\section*{Discussion}
This study investigated the discussion about the novel coronavirus on Twitter. We examined ``fake news''  URLs and misinformation stories spreading in this emergence event. Our study shows that news agencies, government officials, and individual news reporters do send messages that spread widely, and so play critical roles. However, the most influential tweets are those posted by regular users, some of whom are bots. Tweets mentioning ``fake news''  URLs and misinformation stories are more likely to be spread by regular users than the news or government accounts. The distribution of users mentioning the URLs of less credible news sites across countries is different from the distribution of users mentioning real news URLs. More users mentioning these less credible sites and/or the disinformation story-lines come from United States. Unlike messages that mention real news URLs or don't discuss these disinformation story-lines which often spread between countries, these ``fake news'' discussions typically spread within a country. 

In this paper, we utilized machine learning systems to predict users' latent attributes, such as their locations and political orientations. Even though our prediction systems have reasonably high accuracy, they are still prone to prediction errors for individual users. To ensure the reliability of our analysis, we have focused on aggregated results. For the same reason we focused on comparative results instead of absolute values; e.g., there is a higher percentage of conservative users involved in the bio-weapon conspiracy tweets than in non-conspiracy tweets. We also tested the generalizability of the machine learning systems on this COVID-19 dataset. For the location prediction system, we re-trained it using the geotagged users in this dataset. For the identity and political orientation prediction models, we extracted all test users existing in the COVID-19 dataset and re-collected their most recent 200 tweets in this time period. The testing accuracies for this subset of test users are 90.2\% and 91.4\% which are very close to the performance 90.6\% and 90.6\% on identity and political orientation classification respectively evaluated on the original dataset. In order to get a consistent evaluation on the political ideology of global users, we apply a political orientation prediction system trained on users from the United States on English speaking users globally. The political orientation prediction results should be interpreted relative to the current conservative versus liberal differences in the United States. We note that, to first order, these same differences are prevalent in Western Europe.

It is important to note that this study only extracted tweets containing certain types of news URLs or certain keywords associated with disinformation story-lines. Whether these tweets are being spread by those knowing they are inaccurate maliciously, as a joke, or simply to discuss the inaccuracy is not considered in this paper. Future work, separating these tweets by the original purpose could provide us with a better understanding of how disinformation spreads during an emergency and the conditions under which it needs to be countered. Our search criteria for finding tweets about coronavirus resulted in a bias toward tweets in the English language. Our search was also constrained by the limits of the Twitter APIs. Hence, there are likely to be additional conversations related to this pandemic that are not captured. Future work might consider automatically detecting new tracking keywords in the streaming data to dynamically shift the selection and so capture more conversations as the conversation drifts between topics \cite{kumar2019track}. 

There are many potential implications of this study. We found that regular users sent the majority of tweets referring to non-credible news sites and mentioning the disinformation story-lines.  Many of these regular users appear to be bots; however, most people cannot recognize bots. Thus, people should be cautious when reading tweets sent from regular users, and perhaps be even more skeptical when reading those posts, than those from news agencies and the government.  Regular users in some countries appear to be greater consumers of information and sources lacking credibility.  This suggests that local country regulations, policies, and technology may be important in reducing the spread of such information.  We found that health authorities, such as WHO, played a critical role.  Concerted efforts to increase the reach of such authorities may be of value in combating misinformation.

\noindent \textbf{Author contribution}: B. Huang and K. M. Carley conceived the research ideas and experimental design. B. Huang performed data collection and experiments. B. Huang and K. M. Carley analyzed and interpreted results. All the authors wrote and revised the manuscript.

\bibliography{scibib}

\bibliographystyle{Science}


\end{document}